# Mobility of single vacancies and adatoms in graphene at room temperature

*Tuan T. Tran* [1(*)], *Per O. Å. Persson* [2], *Ngan Pham* [3], *Radek Holenak* [1], *and Daniel Primetzhofer* [1]

[1] Department of Physics and Astronomy, Ångström Laboratory, Uppsala University, Box 516, SE-751 20 Uppsala, Sweden

[2] Thin Film Physics Division, Department of Physics, Chemistry, and Biology (IFM), Linköping University, Linköping, Sweden

[3] Division of Solid-State Electronics, Department of Electrical Engineering, Uppsala University, SE-751 03 Uppsala, Sweden

(*) Corresponding author: Tuan T. Tran (tuan.tran@physics.uu.se)

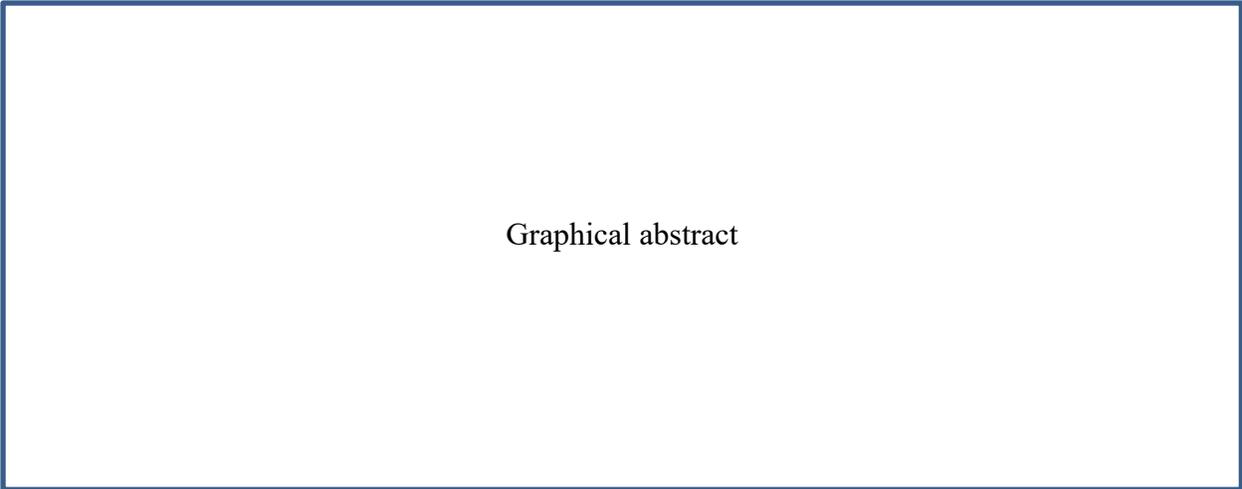

Graphical abstract

**ABSTRACT**

We investigate the mobility of structural defects, adatoms and defect-adatoms combination in self-supporting graphene subjected to keV ion irradiation. In the first scenario, homogeneous irradiation using 20 keV $Ar^+$ ions at a dose of $3\times10^{14}$ ions/cm² induces tensile strain of up to 0.8%. This strain diminishes with increasing defect density at the dose of $5\times10^{14}$ ions/cm², indicating a strain-relaxation mechanism. Contrary to the expected localized behavior, vacancies exhibit long-range interactions, contributing to global strain effects across the lattice. In the second scenario, by employing a nanopore mask, we spatially confined defect generation to periodically aligned circular regions surrounded by non-irradiated material, enabling direct observation of vacancy and adatom dynamics. Selected area electron diffraction (SAED) reveals significant structural damage in areas adjacent to the irradiated regions, suggesting that single vacancies migrate over distances on the order of 100 nm from irradiated to non-irradiated zones even at room temperature. The build-up of lattice strain observed in this study may play a key role in lowering the migration barrier of single vacancies, thereby facilitating their diffusion into pristine lattice regions. Furthermore, the findings highlight the role of pre-existing surface contaminants in preserving lattice integrity through a self-healing mechanism, where adatom-induced lattice reconstruction mitigates defect-induced structural degradation.

**INTRODUCTION**

Graphene, a single layer of carbon atoms arranged in a two-dimensional hexagonal lattice, has garnered immense attention due to its extraordinary physical and chemical properties. However, like all materials, graphene is not immune to structural imperfections. Structural defects, which can arise during synthesis and processing, critically influence the material's overall properties such as electronic, mechanical, thermal, and chemical properties [1,2]. While defects are often seen as detrimental, they can also be leveraged to tailor graphene's properties for specific applications. In some cases, controlled introduction of defects can enhance graphene's chemical reactivity [3,4] or create desirable electronic properties, such as bandgap tuning for semiconductor applications [5,6]. Therefore, understanding the nature, formation mechanism, and effects of structural defects in graphene is crucial for both optimizing its performance in practical applications and developing strategies to either mitigate or harness them.

Structural defects in graphene, which can form naturally or result from irradiation with high-energy electrons and ions, primarily include single vacancies (SVs), double vacancies (DVs), Stone-Wales defects, and line defects (dislocations) [1,2,7]. While most defects in graphene are generally immobile, the migration of SVs, particularly at around room temperature, is debated due to their wide range of reported migration barriers of 0.8–1.8 eV. For instance, molecular dynamics simulations suggest that it would take several years for an SV to migrate beyond a few nanometers with a barrier of 1.4 eV [8]. However, migration via single-atom hopping has been reported with a significantly lower energy barrier of 0.94 eV, reducing migration time to about 10 minutes at room temperature [9]. Another potential migration pathway, involving out-of-plane atomic displacement, is estimated to lower the energy barrier further to 0.56 eV using DFT with local density approximation (LDA) [10]. Therefore, migration of the SVs in graphene might be theoretically possible at room temperature.

Observing the migration of vacancies in graphene is made more complicated due to the self-healing, meaning that the material can spontaneously repair certain structural defects without the need for external stimuli, such as high temperatures or additional chemicals [11–13]. This self-healing process occurs through migration of carbon adatoms, where carbon atoms rearrange themselves to fill vacancies or correct distortions in the lattice. Remarkably, these self-healing processes can sometimes occur at room temperature, driven by the system's natural tendency to minimize energy [13].

In this study, we investigate the migration of structural defects induced by ion irradiation, the diffusion of surface atoms/molecules, and the self-healing process through defect-adatom recombination. Initially, we examine the behavior of the graphene lattice under ion irradiation using a broad beam raster-scanned across the entire graphene sample. Subsequently, we irradiate a similar graphene sample through a suspended nanopore mask, facilitating the observation of defect mobility and adatom/vacancy interactions using transmission electron microscopy (TEM) and selected area electron diffraction (SAED).

**EXPERIMENTS**

Irradiation experiments were performed in two configurations, as illustrated in Fig. 1. In the first configuration, the graphene membranes were irradiated with a 20 keV $Ar^+$ ion beam raster-scanned across the entire sample surface. In the second configuration, the samples were irradiated through a silicon mask containing a pre-fabricated array of circular pores, each 110 nm in diameter, spaced 500 nm apart, and arranged in a 150 × 150 pattern. The mask was fabricated using electron beam lithography and micro-machining techniques, as detailed in Ref. [14].

The irradiation experiments were conducted using a 350 kV Danfysik ion implanter at the Tandem Laboratory, Uppsala University [15]. For the unmasked irradiation, $Ar^+$ ion doses ranged from $1\times10^{14}$ to $5\times10^{14}$ ions/cm². Irradiation of graphene with 20 keV $Ar^+$ ions, modeled using the ZBL potential, leads to atomic displacements whenever the impact parameter is smaller than 0.57 Å, assuming a displacement threshold energy of 22 - 24 eV [16,17]. This results in a damage-sensitive area of approximately 1 Å² per atom. Given that each graphene atom occupies an area of 2.62 Å², the estimated defect density is about 40% of the ion dose. Since graphene has an areal atomic density of $3.8 \times 10^{15}$ atoms/cm², an ion dose of $10^{14}$ ions/cm² corresponds to approximately 1% of carbon atoms being displaced.

In the masked configuration, a single dose of $1\times10^{16}$ ions/cm² was applied, which displaces 100% of carbon atoms according to the previous estimation. All experiments were performed in transmission mode, with the samples mounted over a 2 mm aperture of the sample holder. This setup prevented a potential influence of backscattered and recoiled species that might otherwise occur if the samples were mounted on a substrate.

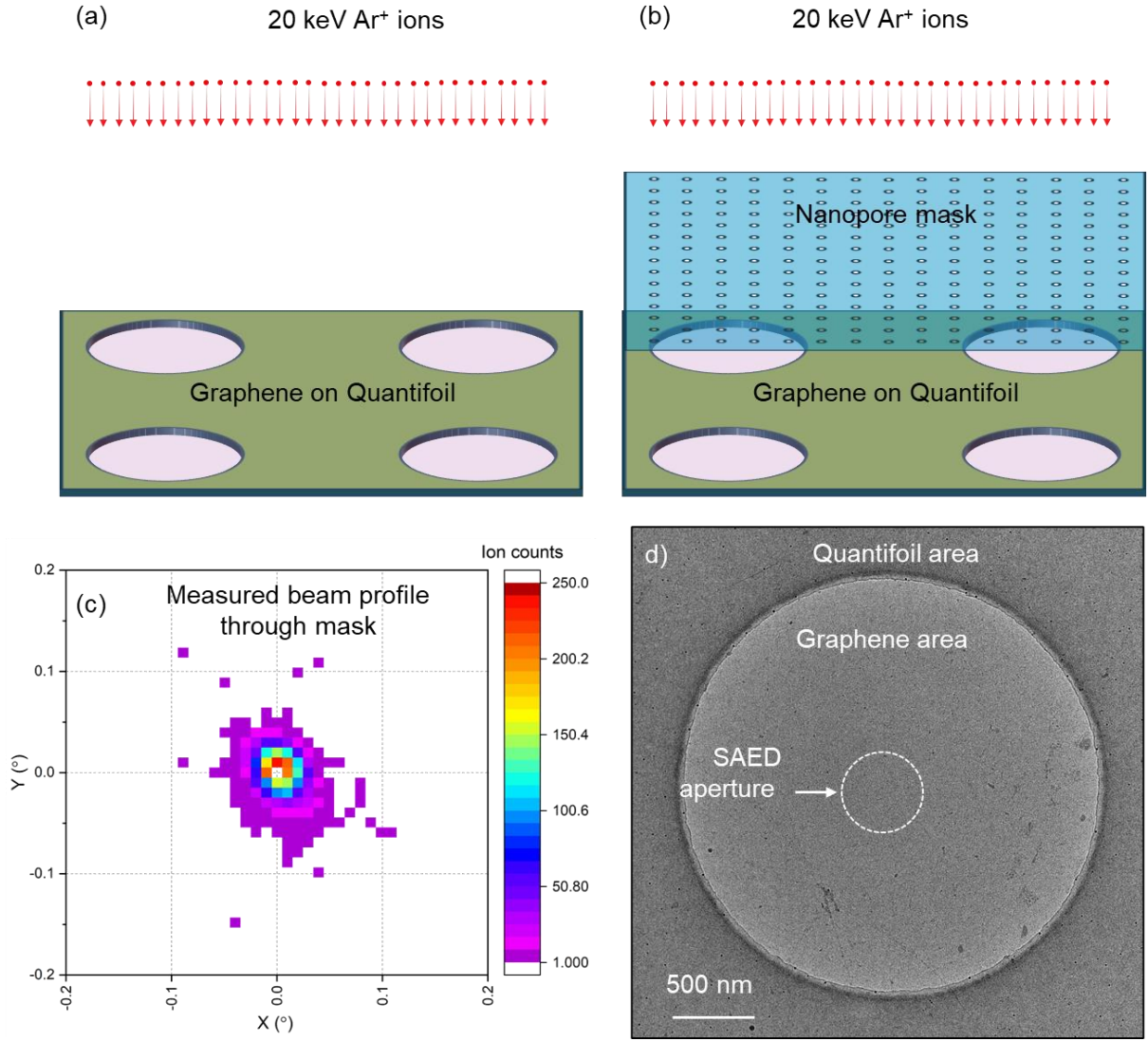

Fig. 1: Self-supporting monolayer graphene irradiated in two different scenarios: with a broad 20 keV Ar$^+$ beam (a), and with the Ar$^+$ beam passing through a nanopore mask (b). Measured intensity profile of the post-mask beam showing an extremely parallel beam with a small divergence of 0.1° (c). A TEM image of the graphene on Quantifoil sample with the SAED aperture indicated by the white dash circle (d).

To evaluate the angular intensity profile of the 20 keV Ar$^+$ ion beam transmitted through the mask, a three-dimensional ion detector within a medium-energy ion scattering (MEIS) chamber was utilized. As described in Ref. [18,19], the detector captures both the spatial distribution and energy of individual ions impacting its surface, enabling the acquisition of a detailed 2D ion beam map.

As shown in Fig. 1c, the post-mask beam exhibits minimal divergence, with 99.5% of its intensity confined within an angular range of ±0.05°. No ion impacts were observed beyond a divergence angle of ±0.1°. These findings confirm the absence of stray ions outside the beam regions defined by the nanopore mask, effectively eliminating the risk of unintended irradiation. The samples used in this study were monolayer graphene mounted on Quantifoil, a thin polymer film with 2.5 µm-diameter circular holes. The graphene membranes, self-supporting within these holes, served as the target areas for irradiation. The samples were purchased from Graphenea. An overview TEM image of a hole in this sample is shown in Fig. 1d. The nanopore mask was placed in direct contact on the samples for the irradiation.

After irradiation, the samples were characterized with an FEI Titan Themis 200 TEM microscope operated at an acceleration voltage of 80 keV at Myfab, Uppsala University. In conventional TEM mode with a broad electron beam, the 80 keV electron energy is sufficiently low to avoid structural damage to the graphene lattice [20]. Our investigation demonstrates that graphene preserves its structural integrity even after 15 minutes of exposure to the 80 keV beam, whereas exposure to a 200 keV beam causes amorphization within seconds. Structural integrity was further evaluated using selected area electron diffraction (SAED) in the TEM system, employing a 40 µm diameter SAED aperture to analyze a ~500 nm area of the membrane. The white dashed circle (500 nm in diameter) in Fig. 1d indicates the area from which electron diffraction patterns (DPs) were obtained.

To quantify surface contamination on graphene, scanning transmission electron microscopy with electron energy loss spectroscopy (STEM-EELS) was employed. Transmitted electrons passing through the graphene membrane were energy-dispersed by an electromagnetic prism (CEFID filter), and the resulting spectrum was recorded at individual pixel positions using an electron-counting detector (Dectris ELA), forming EELS spectra. An EELS map was generated by scanning the electron beam across the sample area. From these spectra, membrane thickness maps, expressed in units of the inelastic mean free path ($\lambda$), were calculated using the formula $t/\lambda = -\ln(I_0/I_t)$, where $I_0$ is the intensity of the zero-loss peak and $I_t$ is the total spectral intensity. The thickness maps were analyzed using the Gatan Digital Micrograph software.

To assess the distribution of defects migrating beyond the irradiated areas, high-resolution transmission electron microscopy (HR-TEM) imaging was performed at progressively outward regions from the irradiation centers. The imaging was conducted using an FEI Titan³ 60-300 TEM microscope at Linköping University, operated at 60 keV with a monochromated electron source and an image-forming Cs-corrector. At this electron energy, the TEM is even less damaging to the graphene lattice than at 80 keV. Under these conditions, the system achieves a spatial resolution of <1.2 Å, sufficient to resolve the graphene lattice and thus enabling FFT-analysis of images.

**RESULTS**

1. *Configuration 1: Graphene irradiated without a nanopore mask.*

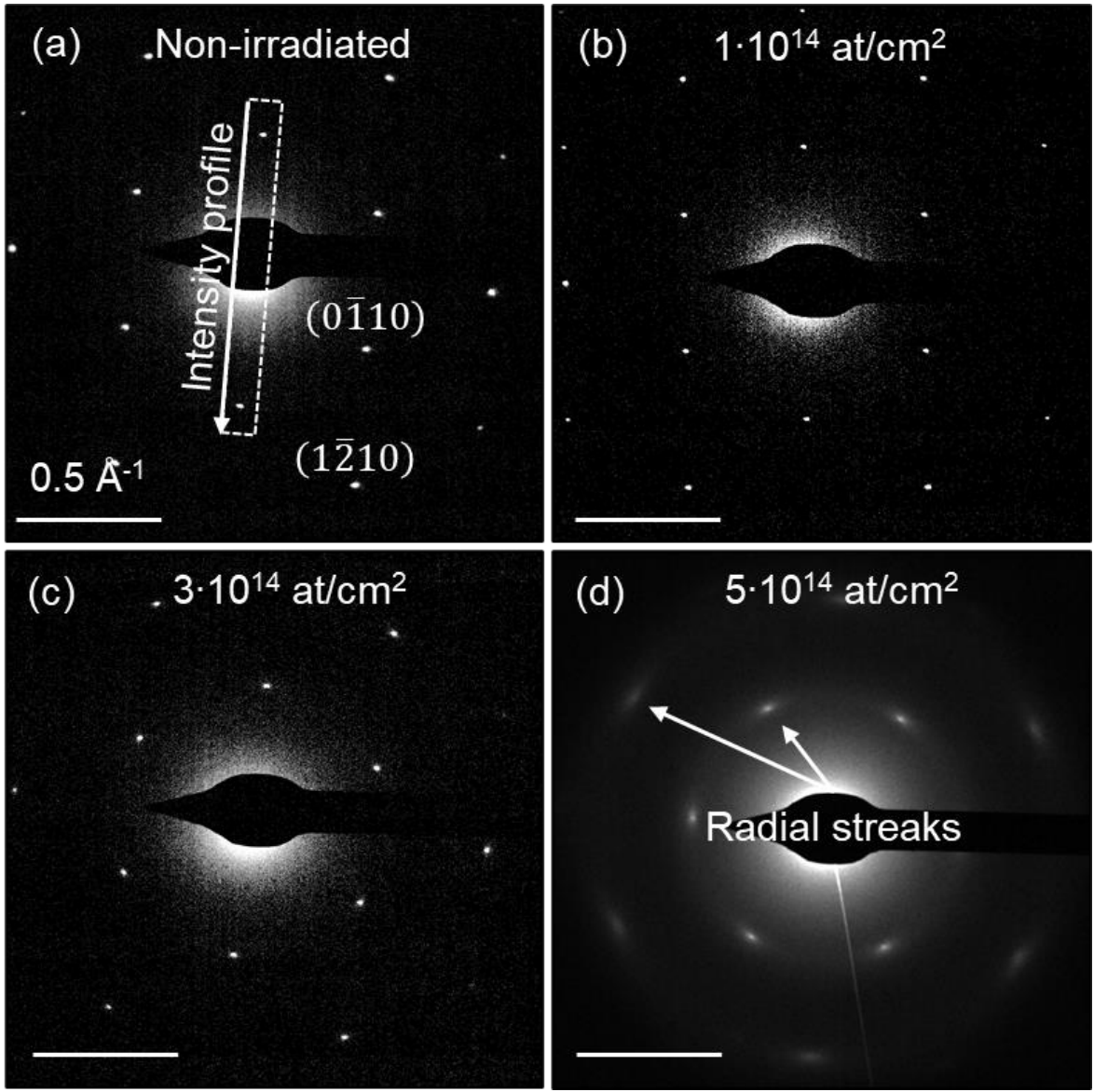

Fig. 2: Electron diffraction patterns of graphene after irradiation with 20 keV Ar$^+$ ions at different doses, 1 - 5×10$^{14}$ at/cm$^2$. In this experiment, no nanoporous mask was employed.

Fig. 2 shows electron diffraction patterns (DPs) of the graphene sample irradiated without a nanopore mask. The pattern of a non-irradiated pristine sample is shown in Fig. 2a, revealing a typical hexagonal pattern of monolayer graphene, with the first- $(0\bar{1}10)$ and the second-order diffraction spots $(1\bar{2}10)$ corresponding to lattice periodicities of 2.13 Å and 1.23 Å, respectively [21].

Up to irradiation doses of $3\times10^{14}$ at/cm$^2$, the graphene retains high structural quality, as evidenced by the sharp, discrete spot diffraction patterns (Fig. 2b, c). However, at a dose of $5\times10^{14}$ at/cm$^2$, the diffracted intensity weakens significantly, indicating lattice deterioration (Fig. 2d). The spots also widen asymmetrically, with a more pronounced spread in the radial direction, which aligns with previous studies where graphene irradiated with 30 keV helium ions at doses between 3 - $12\times10^{15}$ at/cm$^2$ showed similar effects [22]. The radial streak of diffraction spots (Fig. 2d) at higher doses suggests the formation of mosaicity, where the lattice divides into smaller crystal domains with small rotational misalignments.

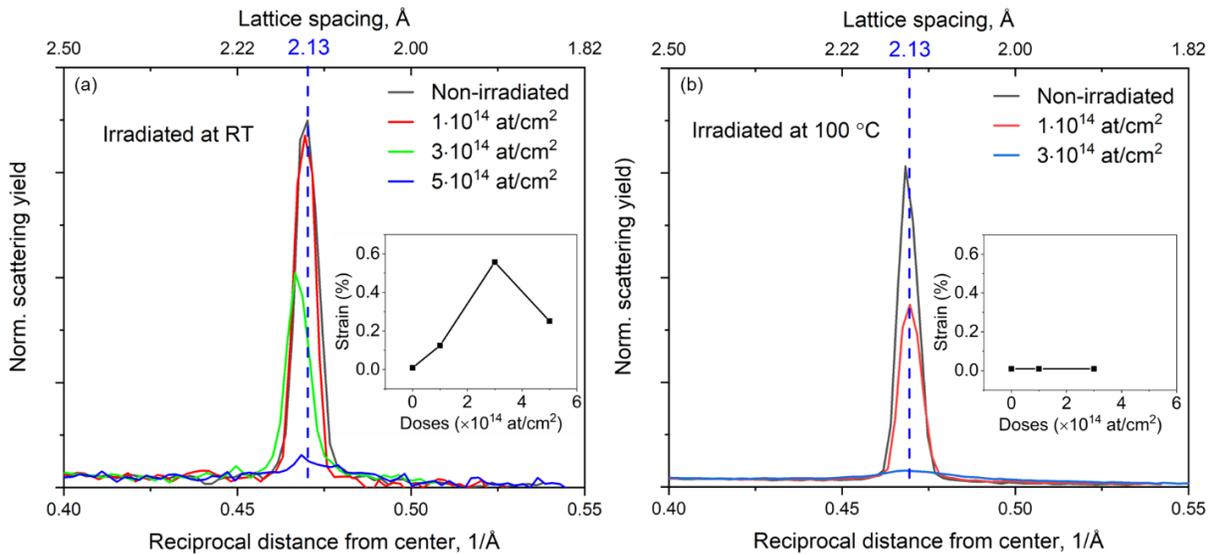

Fig. 3: Profiles of the $(0\bar{1}10)$ diffracted intensity from samples irradiated with different ion doses at room temperature (a) and at 100 °C (b). Insets are strain in the samples as deduced from the shift of the diffracted beams.

Fig. 3 shows the intensity profiles of the diffraction spots for both non-irradiated and irradiated samples. The profiles were obtained by integrating the diffraction intensity over the area indicated in Fig. 2a. The positions of the diffraction peaks were determined by fitting them with Gaussian functions. To minimize errors in peak identification, the two symmetrical peaks were re-centered, and only one peak is displayed in the graph. As shown in Figure 3a, a slight but noticeable shift in peak positions towards smaller reciprocal distances is observed, indicating lattice expansion. The lattice strain reaches a maximum of approximately ~0.8% at an irradiation dose of $3\times10^{14}$ at/cm$^2$. At higher doses, the lattice strain appears to relax, returning closer to its original value.

At 100 °C, the graphene lattice exhibits increased susceptibility to damage, as shown in Fig. 3. Comparable levels of lattice disorder, indicated by similar diffraction intensities, are achieved at significantly lower doses at 100 °C (e.g., $1\times10^{14}$ at/cm$^2$ and $3\times10^{14}$ at/cm$^2$) compared to those required at room temperature ($3\times10^{14}$ at/cm$^2$ and $5\times10^{14}$ at/cm$^2$). Additionally, the inset in Fig. 3b reveals a complete relaxation of lattice strain in graphene irradiated at 100 °C.

2. *Configuration 2: Graphene irradiated with a nanopore mask.*

In this experiment, a similar graphene sample from the same batch was irradiated using a nanopore mask with pore diameters of 110 nm and a pore-to-pore distance of 500 nm. Fig. 4 shows TEM images of the graphene membrane at various locations across the edge of the pattern area, labeled from positions 1 to 5, with precise coordinates indicated in Fig. 4a.

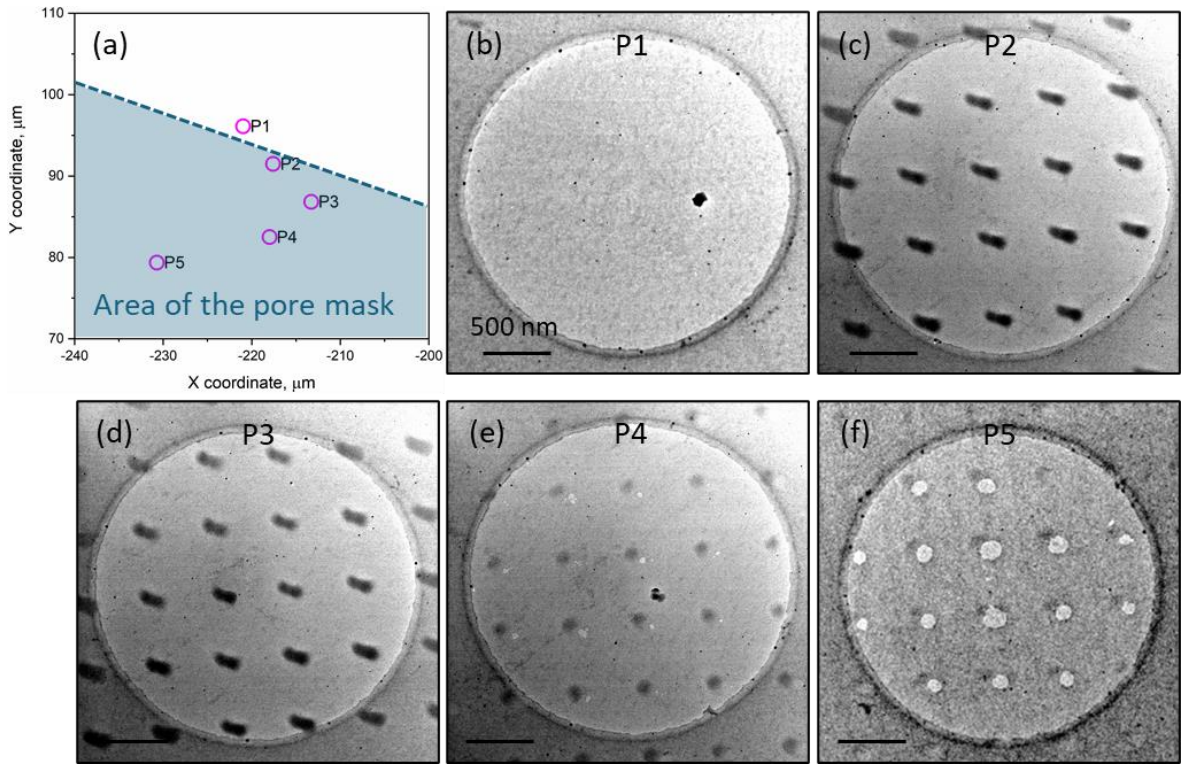

Fig. 4: Coordination of self-supporting graphene on the sample (a). TEM images of graphene at different positions across the border of the irradiating patterns (b – f).

At position 1 (P1), located just outside the pore area, no distinct structures are observed. In contrast, at P2, immediately across the edge of the patterning area, the graphene membrane exhibits elongated dark spots. These patterns arise from the ion beam passing through the circular pores of the mask, with elongations measuring less than 100 nm, likely due to thermal drift between the mask and the sample during irradiation. In conventional TEM mode, these dark spots correspond to thicker regions, indicating an accumulation of contaminants under ion beam exposure. Surface contamination in graphene primarily originates from the transfer process. During the transfer of graphene from its growth substrate onto a target surface, remnants of the support layer are difficult to fully remove, even with extensive cleaning [23,23,24].

Farther from the edge, at P4, the graphene membrane displays perforations with pore sizes ranging from 20 to 50 nm. The largest pores, approximately 110 nm in diameter, are observed at position 5, the farthest point from the border.

Although the entire irradiated area was exposed to the same ion fluence and flux, the resulting patterns in graphene varied, evolving from the buildup of contamination at P2 and P3 to membrane perforation at P4 and P5. This variation arises from the combined effects of electronic and nuclear interactions between the irradiating ions and the graphene, as well as the resulting inhomogeneous concentration of mobile contaminants near the edge of the porous mask. A discussion of this observation has been presented in details in Ref. [14].

To characterize the distribution of contaminants on graphene, we quantified the membrane thickness in units of inelastic mean free path $\lambda$ using STEM-EELS. Fig. 5a-d show the mean thickness distributions of the membrane at positions 2–5, respectively. Fig. 5e, derived from integrating the mean thickness within profile window 1, reveals that position 2 exhibits a peak with a maximum thickness of approximately ~0.35 $\lambda$, indicating contaminant accumulation. At P3, the thickness of this peak decreases to ~0.2 $\lambda$. At P4, the thickness profile reveals a valley approximately 30 nm in diameter, indicating graphene perforation. By position 5, this perforation expands to approximately 110 nm. This progression highlights a clear structural pattern: as the distance from the border increases, the features evolve from contaminant accumulation to small pores and eventually to larger perforations.

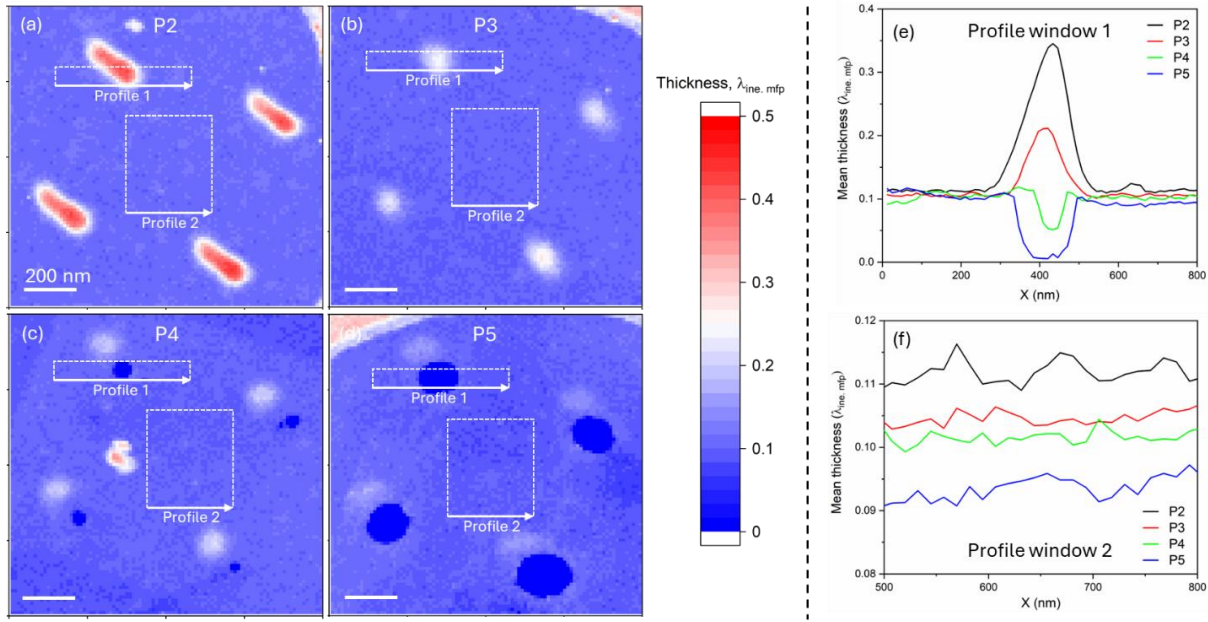

Fig. 5: Thickness distribution, in unit of inelastic mean free path λ, of the sample at different positions after irradiation with the mask (a-d). Thickness profiles within window 1 and 2 of the maps (e, f).

In parallel with this structural evolution, the level of residual contamination at the center of the four exposed areas decreases, as shown in Fig. 5f. This observation supports the hypothesis that mobile contaminants diffuse into the exposed areas and are progressively removed by the ion beams. The eventual depletion of contaminants leads to the perforation of the membranes.

To assess the crystal quality of the patterned membranes, selected area electron diffraction (SAED) was utilized. Fig. 6a presents a TEM image of graphene at P5, featuring large pores at the four corners. The white dashed circle indicates the selected area for electron diffraction, with the entire diffraction region situated outside the irradiated areas. At P1, located ~2 μm outside the irradiated pattern, far enough that no defects can diffuse there, the SAED pattern exhibits sharp, spot-like diffraction, indicating a high-quality crystal. This spot-like appearance is also maintained at P2 and P3, which are located within the area of the pore mask and ~2 – 5 μm from the edge of the pattern. As shown in Fig. 4, the patterns observed at these positions correspond to the membrane with contamination buildup. At P4, where the membrane has been perforated with small pores, the SAED reveals an evident decline in crystal quality. Lastly, at P5, the SAED indicates significant structural defects in the graphene, corresponding to the formation of large pores after irradiation.

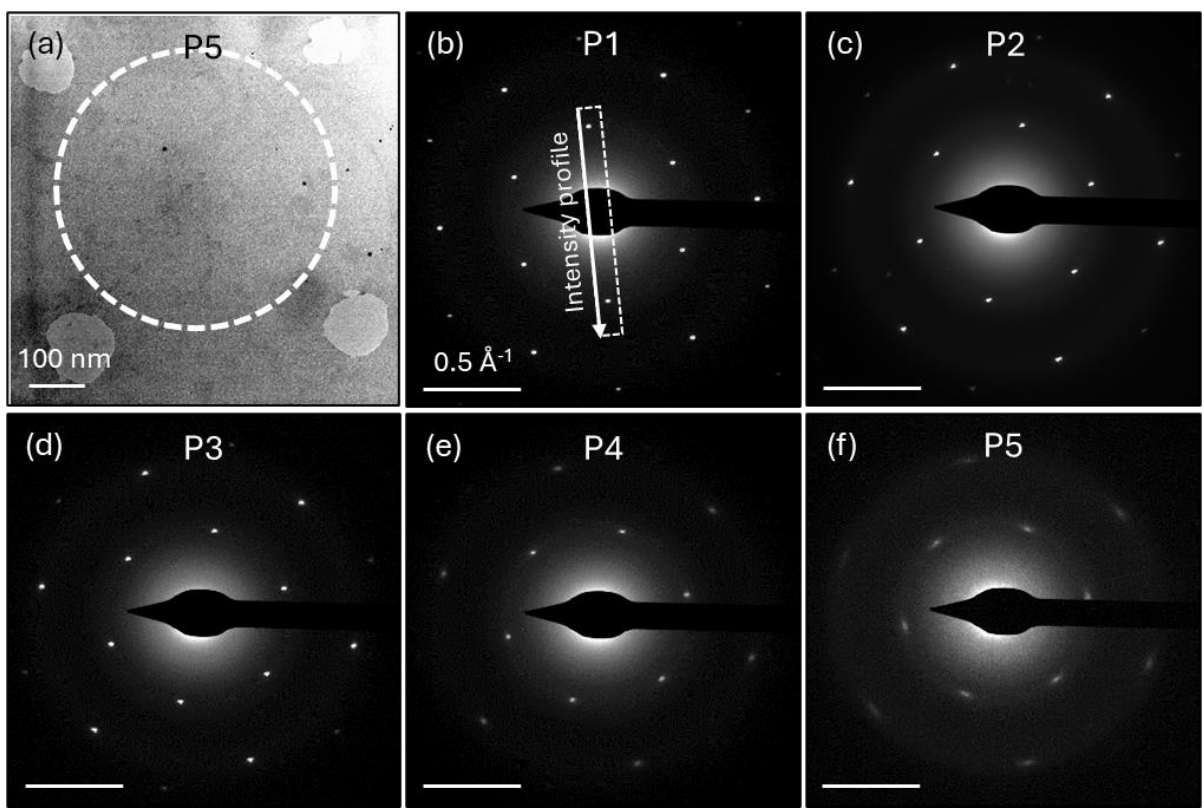

Fig. 6: TEM image of graphene at position P5 on the sample (a). Selected area of electron diffraction indicated with the dash circle. SAED patterns of non-irradiated graphene within the irradiating patterns (b - f).

Fig. 7 presents the intensity profiles of the diffraction spots, obtained by integrating the diffraction signals within the white dashed windows represented in Fig. 6b. The most intense peak corresponds to the membrane at P1, which lies outside the mask's area. Although the diffraction patterns at P2 and P3 remain of high quality (as seen in Fig. 6), the intensity profiles of the diffraction spots at these positions are only half as strong as those at P1. A more substantial reduction in diffraction intensity is observed at positions 4 and 5, where the membrane is perforated with small and large pores, respectively.

The reduced diffraction intensity corresponds to a shift in the peak centers towards smaller reciprocal distances for the P2 and P3 curves, indicating global tensile strain in the lattice at these positions. By fitting the peaks with a Gaussian function, the peak centers—and consequently the lattice strain—were determined, as shown in the inset of Fig. 7. A tensile strain of up to 0.8% was observed at position 2. At higher defect levels, the strain in the membrane at positions 4 and 5 relaxes to nearly zero. A comparison of Fig. 3a (non-masked sample) and Fig. 7 (masked sample) reveals similar defect and strain characteristics in the graphene membranes. This similarity persists despite the non-masked membrane being directly exposed to the ion beam, whereas the masked sample's membrane was shielded by the mask.

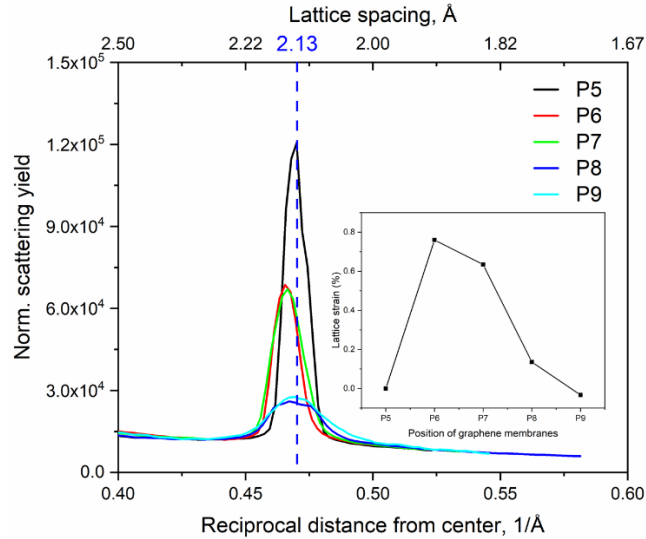

Fig. 7: Intensity profiles of diffracted electron beams at different non-irradiated positions across the border of the irradiating patterns.

To assess defect migration beyond the irradiated areas, HR-TEM imaging combined with Fourier transform analysis was conducted from the edge of irradiated regions to the center of the masked regions, as shown in Fig. 8. The dashed circles in Fig. 8a indicate the 110 nm diameter irradiated regions, while the small squares mark the locations where sequential HR-TEM images were captured. Fig. 8b presents an HR-TE; image taken at point 5, located at the center of the masked region, with its inset displaying the fast Fourier transform (FFT). The FFT confirms the crystallinity of graphene, as evidenced by the six-fold symmetry of reflections.

The defect density at various points was evaluated by comparing the intensity of these spots. Fig. 8c shows the intensity of the diffraction spots radially integrated from the FFTs, revealing a clear defect gradient—regions closer to the irradiated areas exhibit higher defect densities than those farther away. This trend further supports the notion that defects migrate from the irradiated areas into the surrounding regions.

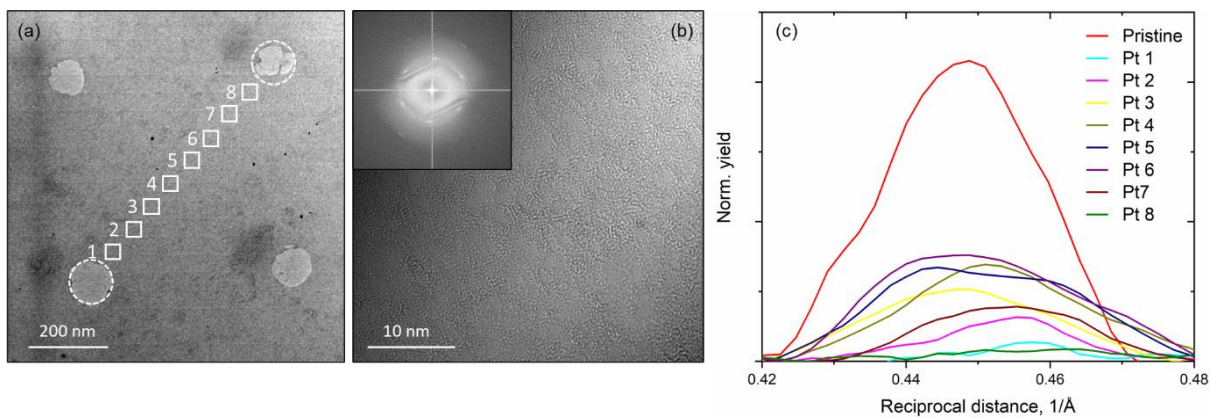

Fig. 8. Structural integrity in the vicinity of the irradiated regions. (a) TEM images of graphene after irradiation. (b) A HR-TEM image taken at point 5, inset is the FFT of the HR-TEM image. (c) Radially integrated intensity of the reflection spots from the FFT images at positions marked in (a).

**DISCUSSION**

1. *Lattice strain in ion irradiated graphene*

Defects in graphene, such as vacancies and dislocations, generate localized strain due to the redistribution of bond forces in their vicinity. This strain arises because the removal or misalignment of atoms alters the equilibrium bonding environment, leading to compressive or tensile stresses in the surrounding lattice. Studies have shown that the strain field around these defects reaches far into the unperturbed hexagonal network [25] and induce long-range strain in the lattice [26]. This phenomenon is particularly relevant in defect-engineered graphene applications, where controlled strain manipulation can be utilized to tailor material properties for specific functionalities [27].

Theoretical studies have also demonstrated that strain fields induced by multi-vacancy defects in graphene strongly influence the diffusion of single vacancies, ultimately promoting the formation of extended line defects [28]. Using density functional theory (DFT) and kinetic Monte Carlo simulations, researchers have mapped out the potential energy surface for single-vacancy migration near vacancy clusters. Their findings reveal that strain anisotropically modifies the activation energy barriers for vacancy diffusion, significantly biasing the direction of vacancy migration. As a result, vacancies tend to coalesce into linear complexes aligned along the primary crystallographic directions, rather than forming compact clusters. These extended defects may further evolve into dislocations or grain boundaries.

Experiments using *in situ, in operando*, reflection high-energy electron diffraction have demonstrated that global tensile strains in graphene can originate from localized strains around point defects, which are generated during processes like $O_2$ plasma etching at high temperatures or ion bombardment [29,30].

In this study, we present direct evidence of global lattice expansion of up to 0.8% in self-supporting graphene induced by ion irradiation. In the self-supporting case, strain appears to relax at higher ion doses, contrasting with the behavior observed in graphene on an Ir(111) substrate [29]. This discrepancy likely arises from the more complex substrate interactions in the graphene/Ir(111) system, which inhibit strain relaxation and allow for the accumulation of higher tensile strains. Due to its simpler structure, our study provides a well-defined system that can improve theoretical models and enhance our understanding of strain induced by defects in graphene.

2. *Migration of vacancies in graphene at room temperature*

The SAED data of graphene irradiated with the nanopore mask in Fig. 7 reveal that the membrane exhibits defects even in areas supposedly outside the irradiated region. Stray ions, which could potentially cause these unexpected defects, can be ruled out. Our 3D-MEIS measurements of the ion beam intensity passing through the mask confirm an almost perfectly parallel beam, with 100% of the post-mask intensity confined within a divergence angle of 0.1°, as shown in Fig. 1c. Backscattered and recoiled ions can also be excluded as a cause, given that the graphene was irradiated in transmission mode, with the sample mounted on a perforated holder. While transmitted ions could theoretically backscatter from the backside of the implanter chamber, our SRIM-based estimation with the employed geometry indicates that an ion has a probability lower than $10^{-6}$ to backscatter toward the sensitive area of the sample. These observations suggest that the lattice imperfections observed outside the irradiated areas result from extended defect migration within graphene. This conclusion is further reinforced by the data in Fig. 8, which shows a clear gradient in defect density, indicating that defects propagate from the irradiated regions into the surrounding lattice.

The thermodynamic behavior of single vacancies at room temperature remains a subject of debate. While high migration barriers (1.3–1.7 eV) [8,31,32] suggest that single vacancies should be immobile at RT, alternative studies propose significantly lower barriers (~0.87–0.94 eV) [9] [10], making vacancy migration feasible on much shorter timescales. The potential role of out-of-plane atomic displacements in reducing activation energy further supports this possibility, aligning with DFT calculations. These discrepancies highlight the need for further experimental validation to clarify the true mobility of single vacancies at RT.

Experimentally, there is evidence for the migration of vacancies in graphene at RT. Using high-resolution TEM, Hashimoto et al. observed vacancies as white dots shifting positions between two frames taken within a few hundred seconds [33]. Similarly, Kotakoski et al. employed atomic-resolution scanning TEM, showing random walks of point defects in graphene at RT [34]. However, the authors suggested that the observed defect migration was not solely due to thermal excitation, but rather caused by collisions between imaging electrons and target nuclei. While direct imaging of vacancy migration using TEM techniques provides atomic-level insights, the interaction between high-energy electrons and the graphene lattice complicates attributing this behavior solely to thermal excitation.

In this study, using the nanopore mask for irradiation the areas where the defects are generated and imagined are decoupled. Since the TEM imaging with electron energy of 80 keV does not induce defects in graphene. The fact that structural defects were observed outside the irradiated regions suggests that the detected defects in the non-irradiated regions migrated spontaneously from irradiated zones.

The migration of single vacancies in graphene at RT can be attributed to several factors that lower the activation energy for diffusion, enabling significant defect mobility. One key factor contributing to this lower barrier is the presence of tensile strain in the graphene lattice. Strain modifies the bonding environment, reducing the energy required for vacancy hopping between adjacent lattice sites, making diffusion feasible at lower temperatures than previously expected.

Additionally, the interaction between single vacancies and complex multi-vacancy structures can plays a crucial role in vacancy migration and defect evolution. As single-vacancies coalesce into larger vacancy clusters, they generate significant strain fields that influence the diffusion pathways of nearby single vacancies. This strain not only reduces the activation energy for migration but also guides vacancies into forming extended line defects [28]. These line defects might propagate deep into the surrounding lattice, disrupting the otherwise pristine graphene structure. The formation of these extended defects effectively divides the graphene sheet into smaller lattice domains with slight misorientations. This suggestion is evidenced by the selected area electron diffraction (SAED) patterns observed in Fig. 2 and Fig. 6.

3. *Migration of surface contaminants and self-healing of graphene's lattice*

In addition to the degradation of crystal quality in non-irradiated regions, the defect density varies significantly across different positions on the membranes, as shown in Fig. 6. Notably, the crystal quality tends to decline with increasing distance from the edge of the pore mask toward the center of the irradiating pattern. According to Fig. 5f and Fig. 6, the crystal quality is closely correlated with the level of residual surface contamination. Regions with higher levels of contamination, such as P2 and P3, exhibit comparatively less lattice degradation. Their diffraction patterns retain the spot-like appearance. In regions where small (P4) and large pores (P5) have formed, due to the depletion of residual contamination, crystal quality shows pronounced deterioration.

This observation can be attributed to the spontaneous self-healing property of defective graphene. When defects such as vacancies or cracks form in graphene, carbon atoms from nearby areas or external sources can migrate to fill these gaps [11–13,35]. Studies have demonstrated that graphene can efficiently heal defects over time, even at room temperature [13,35].

At higher temperatures, our data indicate that the overall structural quality of graphene is more damaged after irradiation at 100 °C than at RT as shown by Fig. 3. Temperature plays a crucial role in the creation and behavior of defects in graphene, influencing both the formation rate [36], diffusion of these defects [8,9], the desorption of the contamination [24] and the self-healing process [13,35].

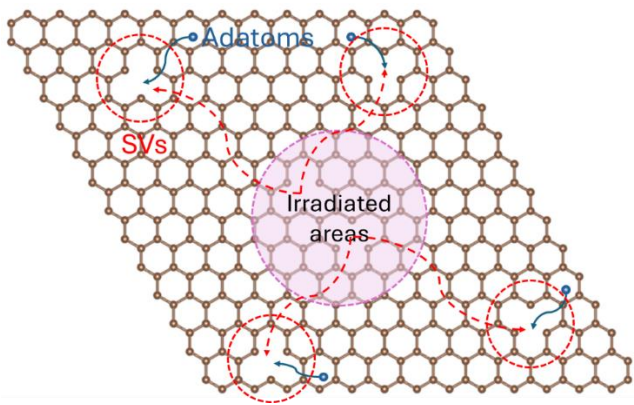

Fig. 9: The dynamics of structural defects and adatoms of graphene at ambient temperature.

The majority of surface contaminants appear to stem from the synthesis and transfer processes of graphene rather than environmental exposure. This suggestion is supported by the lack of accumulation of contaminants on samples irradiated with the nanopore mask, when they were examined with SEM even after weeks of exposure to ambient conditions.

This study uncovers a complex interaction between structural defects, adatoms, and their recombination dynamics, as illustrated in Fig. 9. Ion irradiation generates both single and complex vacancies in the irradiated regions. While complex vacancies remain immobile, single vacancies exhibit mobility and can migrate beyond the irradiated zones at ambient temperature. Additionally, surface contaminants demonstrate high mobility at room temperature. These mobile contaminants play a crucial role in combining the migrating vacancies, aiding in the repair of the crystal lattice.

**CONCLUSION AND OUTLOOK**

This study presents a detailed investigation of structural defects in graphene induced by high-energy ion irradiation, focusing on their mobility and interactions with adatoms. Without the use of a mask, ion irradiation generates tensile strain of up to 0.8% at a dose of $3 \times 10^{14}$ ions/cm². However, as defect density increases, this strain gradually diminishes, indicating strain relaxation due to defect accumulation. The findings also reveal long-range defect interactions, leading to global strain effects across the lattice.

By utilizing a nanopore mask, we successfully isolated defect-generating regions, allowing for a precise analysis of vacancy and adatom dynamics. Selected area electron diffraction (SAED) reveals significant structural damage extending beyond the directly irradiated regions, suggesting that single vacancies migrate from irradiated to non-irradiated areas at room temperature. This observation implies that strain within the graphene lattice may reduce the migration barrier, facilitating vacancy diffusion deep into the unperturbed regions. Furthermore, the study underscores the crucial role of surface contaminants in preserving lattice integrity through spontaneous self-healing processes, which may mitigate defect-induced degradation.

**AUTHOR INFORMATION**

**Corresponding Author**

Tuan T. Tran, Department of Physics and Astronomy, Ångström Laboratory, Uppsala University, Box 516, SE-751 20 Uppsala, Sweden

Email address: Tuan.Tran@physics.uu.se

ORCID: 0000-0002-1393-1723

**Funding Sources**

- The Swedish Research Council VR-RFI (2019-00191).

- The ÅFORSK FOUNDATION (Ref: 22-313)

- The I BERGHS FOUNDATION.

**ACKNOWLEDGEMENT**

T.T.T. gratefully acknowledges financial support from the I. Berghs Foundation and a Young Researcher Grant from the Åforsk Foundation (grant number 22-313). We acknowledge the Tandem Laboratory, a national infrastructure at Uppsala University, for the access to the ion irradiation experiments. Accelerator operation is supported by the Swedish Research Council (grant #2019-00191). We acknowledge Myfab Uppsala for providing facilities and experimental support. Myfab is funded by the Swedish Research Council (2019-00207) as a national research infrastructure. We further acknowledge the Swedish Research Council and Swedish Foundation for Strategic Research acknowledged for access to ARTEMI, the Swedish National Infrastructure in Advanced Electron Microscopy (2021-00171 and RIF21-0026).

**DECLARATION OF COMPETING INTERESTS**

The Authors declare no Competing Financial or Non-Financial Interests.

**DATA AVAILABILITY**

The datasets generated during and/or analysed during the current study are available from the corresponding author on reasonable request.

**REFERENCES**

1. Xu, T. & Sun, L. 5 - Structural defects in graphene. in *Defects in Advanced Electronic Materials and Novel Low Dimensional Structures* (eds. Stehr, J., Buyanova, I. & Chen, W.) 137–160 (Woodhead Publishing, 2018). doi:10.1016/B978-0-08-102053-1.00005-3.

2. Banhart, F., Kotakoski, J. & Krasheninnikov, A. V. Structural Defects in Graphene. *ACS Nano* **5**, 26–41 (2011).

3. Bhatt, M. D., Kim, H. & Kim, G. Various defects in graphene: a review. *RSC Adv.* **12**, 21520–21547 (2022).

4. Boukhvalov, D. W. & Katsnelson, M. I. Chemical Functionalization of Graphene with Defects. *Nano Lett.* **8**, 4373–4379 (2008).

5. Bai, J., Zhong, X., Jiang, S., Huang, Y. & Duan, X. Graphene nanomesh. *Nat. Nanotechnol.* **5**, 190–194 (2010).

6. Lahiri, J., Lin, Y., Bozkurt, P., Oleynik, I. I. & Batzill, M. An extended defect in graphene as a metallic wire. *Nat. Nanotechnol.* **5**, 326–329 (2010).

7. Telkhozhayeva, M. & Girshevitz, O. Roadmap toward Controlled Ion Beam-Induced Defects in 2D Materials. *Adv. Funct. Mater.* **n/a**, 2404615.

8. Dyck, O. *et al.* The role of temperature on defect diffusion and nanoscale patterning in graphene. *Carbon* **201**, 212–221 (2023).

9. Lee, G.-D. *et al.* Diffusion, Coalescence, and Reconstruction of Vacancy Defects in Graphene Layers. *Phys. Rev. Lett.* **95**, 205501 (2005).

10. Wadey, J. D. *et al.* Mechanisms of monovacancy diffusion in graphene. *Chem. Phys. Lett.* **648**, 161–165 (2016).

11. Chen, J. *et al.* Self healing of defected graphene. *Appl. Phys. Lett.* **102**, 103107 (2013).

12. VijayaSekhar, K., Acharyya, S. G., Debroy, S., Miriyala, V. P. K. & Acharyya, A. Self-healing phenomena of graphene: potential and applications. *Open Phys.* **14**, 364–370 (2016).

13. Zan, R., Ramasse, Q. M., Bangert, U. & Novoselov, K. S. Graphene Reknits Its Holes. *Nano Lett.* **12**, 3936–3940 (2012).

14. Tran, T. T., Bruce, H., Pham, N. H. & Primetzhofer, D. A contactless single-step process for simultaneous nanoscale patterning and cleaning of large-area graphene. *2D Mater.* **10**, 025017 (2023).

15. Ström, P. & Primetzhofer, D. Ion beam tools for nondestructive in-situ and in-operando composition analysis and modification of materials at the Tandem Laboratory in Uppsala. *J. Instrum.* **17**, P04011 (2022).

16. Banhart, F. The Formation and Transformation of Low-Dimensional Carbon Nanomaterials by Electron Irradiation. *Small* **n/a**, 2310462.

17. Lehtinen, O. *et al.* Effects of ion bombardment on a two-dimensional target: Atomistic simulations of graphene irradiation. *Phys. Rev. B* **81**, 153401 (2010).

18. Tran, T. T., Jablonka, L., Lavoie, C., Zhang, Z. & Primetzhofer, D. In-situ characterization of ultrathin nickel silicides using 3D medium-energy ion scattering. *Sci. Rep.* **10**, 10249 (2020).

19. Tran, T. T., Lavoie, C., Zhang, Z. & Primetzhofer, D. In-situ nanoscale characterization of composition and structure during formation of ultrathin nickel silicide. *Appl. Surf. Sci.* **536**, 147781 (2021).

20. Börrnert, F. *et al.* Amorphous Carbon under 80 kV Electron Irradiation: A Means to Make or Break Graphene. *Adv. Mater.* **24**, 5630–5635 (2012).

21. Meyer, J. C. *et al.* The structure of suspended graphene sheets. *Nature* **446**, 60–63 (2007).

22. Pan, C.-T. *et al.* In-situ observation and atomic resolution imaging of the ion irradiation induced amorphisation of graphene. *Sci. Rep.* **4**, 6334 (2014).

23. Lin, Y.-C. *et al.* Graphene Annealing: How Clean Can It Be? *Nano Lett.* **12**, 414–419 (2012).

24. Tripathi, M. *et al.* Cleaning graphene: Comparing heat treatments in air and in vacuum. *Phys. Status Solidi RRL – Rapid Res. Lett.* **11**, 1700124 (2017).

25. Cretu, O. *et al.* Migration and Localization of Metal Atoms on Strained Graphene. *Phys. Rev. Lett.* **105**, 196102 (2010).

26. Kotakoski, J., Krasheninnikov, A. V. & Nordlund, K. Energetics, structure, and long-range interaction of vacancy-type defects in carbon nanotubes: Atomistic simulations. *Phys. Rev. B* **74**, 245420 (2006).

27. Qi, Y. *et al.* Recent Progress in Strain Engineering on Van der Waals 2D Materials: Tunable Electrical, Electrochemical, Magnetic, and Optical Properties. *Adv. Mater.* **35**, 2205714 (2023).

28. Trevethan, T., Latham, C. D., Heggie, M. I., Briddon, P. R. & Rayson, M. J. Vacancy diffusion and coalescence in graphene directed by defect strain fields. *Nanoscale* **6**, 2978–2986 (2014).

29. Blanc, N., Jean, F., Krasheninnikov, A. V., Renaud, G. & Coraux, J. Strains Induced by Point Defects in Graphene on a Metal. *Phys. Rev. Lett.* **111**, 085501 (2013).

30. Robinson, J. T. *et al.* Graphene Strained by Defects. *ACS Nano* **11**, 4745–4752 (2017).

31. Krasheninnikov, A. V., Lehtinen, P. O., Foster, A. S. & Nieminen, R. M. Bending the rules: Contrasting vacancy energetics and migration in graphite and carbon nanotubes. *Chem. Phys. Lett.* **418**, 132–136 (2006).

32. El-Barbary, A. A., Telling, R. H., Ewels, C. P., Heggie, M. I. & Briddon, P. R. Structure and energetics of the vacancy in graphite. *Phys. Rev. B* **68**, 144107 (2003).

33. Hashimoto, A., Suenaga, K., Gloter, A., Urita, K. & Iijima, S. Direct evidence for atomic defects in graphene layers. *Nature* **430**, 870–873 (2004).

34. Kotakoski, J., Mangler, C. & Meyer, J. C. Imaging atomic-level random walk of a point defect in graphene. *Nat. Commun.* **5**, 3991 (2014).

35. Vinchon, P., Hamaguchi, S., Roorda, S., Schiettekatte, F. & Stafford, L. Self-healing kinetics in monolayer graphene following very low energy ion irradiation. *Carbon* **233**, 119852 (2025).

36. Chirita Mihaila, A. I., Susi, T. & Kotakoski, J. Influence of temperature on the displacement threshold energy in graphene. *Sci. Rep.* **9**, 12981 (2019).